# Integrated collinear refractive index sensor with Ge PIN photodiodes

Lion Augel*[1], Yuma Kawaguchi[2], Stefan Bechler[1], Roman Körner[1], Jörg Schulze[1], Hironaga Uchida[2], Inga A. Fischer[1]

[1]Institute of Semiconductor Engineering, University of Stuttgart, Stuttgart, Germany

[2]Department of Electrical and Electronic Information Engineering, Toyohashi University of Technology, Toyohashi, Japan

**Abstract**

Refractive index sensing is a highly sensitive and label-free detection method for molecular binding events. Commercial implementations of biosensing concepts based on plasmon resonances typically require significant external instrumentation such as microscopes and spectrometers. Few concepts exist that are based on direct integration of plasmonic nanostructures with optoelectronic devices for on-chip integration. Here, we present a CMOS-compatible refractive index sensor consisting of a Ge heterostructure PIN diode in combination with a plasmonic nanohole array structured directly into the diode Al contact metallization. In our devices, the photocurrent can be used to detect surface refractive index changes under simple top illumination and without the aid of signal amplification circuitry. Our devices exhibit large sensitivities > 1000 nm per refractive index unit in bulk refractive index sensing and could serve as prototypes to leverage the cost-effectiveness of the CMOS platform for ultra-compact, low-cost biosensors.

**Introduction**

On-chip biosensors can serve as diagnostic tools for a wide range of applications that benefit from a combination of high sensitivity and miniaturization. Sensors for the detection of gas molecules and biochemical substances have the potential to drive personalized healthcare solutions that significantly improve quality of life[1], to enable monitoring in industrial processes[2], to improve food safety via the detection of food pathogens[3] and to facilitate environmental monitoring for long-term sustainability[4]. They can be used for self-monitoring and point-of-care solutions in emergency medicine as well as to provide healthcare solutions for an aging society. Biosensors that are fast and simple to operate can make bioanalysis accessible to a broader field of applications and spectrum of users.

While the biosensor market is still dominated by electrochemical sensors, optical sensors have the advantage that they exhibit high sensitivities, are compact and can often be fabricated at low cost. One of the main sensing principles used for optical sensors is refractive index (RI) sensing, where the presence of the target molecule leads to a detectable modification of the refractive index close to the sensor surface. Refractive index biosensors can e.g. be realized using Mach-Zehnder-Interferometers[5,6], ring oscillators[7–9], photonic crystals[10] or plasmonic nanostructures, whose optical response is particularly sensitive to RI changes induced by biological and chemical substances in the immediate vicinity[11–13]. Plasmonic nanohole arrays (NHA) fabricated in thin metal films (typically Au or Al) on a dielectric substrate have been shown to attain high sensitivities in RI sensing in liquids[14,15].

While biosensors using surface plasmon resonances have been commercialized,[16] they require significant and bulky external instrumentation such as microscopes and spectrometers, which limits their applicability. At the same time, plasmonic nanostructures have the potential for further miniaturization. Fabricating such sensors on a Si-based platform using complementary metal-oxide-semiconductor (CMOS) technology could make them available to a broad markets in personalized medicine and environmental monitoring through reduced costs in high volume fabrication[17]. Additionally, the CMOS platform enables the integration of biosensors with signal processing and communications circuitry in a single chip.

To date, there is only a limited number of studies on the integration of plasmonic nanostructures with CMOS-compatible devices[13,18]. The potential of surface illuminated plasmonic sensing setups using an array of nanoapertures in a thin gold film has been shown in devices with separated detector and transducer[19]. The metallic film acts as a RI detector which modulates the spectral transmission through the film depending on the RI in its proximity. Further integration by combining the plasmonic detector and the transducer within a single device can lead to a decrease in sensitivity[18,20] making external instrumentation such as lock-in amplifiers again inevitable.

**Device design**

In this study, we investigate a vertical Ge PIN photodiode with Al NHAs for RI sensing[21] (Fig. 1a). The integration of Ge in a CMOS process is already state-of-the-art and allows to achieve high photocurrents as a consequence of the higher absorption coefficient of the material compared to Si. Furthermore, Ge photodetectors are able to address wavelengths up to ~1.55 µm, thus enabling measurements in a wavelength range where absorption in fluids such as blood is reduced. In our compact device, Al NHAs are structured directly into the photodiode contact metallization. We were able to detect bulk RI changes directly as electrical currents, achieving sensitivities above 1000 nm per RI unit, as an important step towards fabricating highly sensitive biosensors fully integrated on-chip.

The semiconductor layer stack of our device consists of a Ge layer sandwiched between doped Si layers epitaxially grown on a Si wafer. The NHAs are separated from the semiconductor layers by a $SiO_2$ layer with a thickness of $t_{SiO2}$ = 50 nm and have a square lattice geometry. Three parameters (Ge layer thickness $t_{Ge}$, see Fig. 1b, nanohole diameter $d_{NH}$ and the pitch of the arrays $\Lambda$, see Fig. 1c) were varied in order to study their influence on the device responsivity and sensitivity; Fig. 1d gives an overview of the parameters used within the experiment. All samples were characterized experimentally by measuring responsivities in the presence of different superstrate refractive indices. This was achieved by submerging the devices in liquids with different refractive indices – Deionized Water (DI, $n_{DI}$ =1.320[22]), Ethanol (EtOH, $n_{EtOH}$ = 1.353[23]) and Isopropanol (IPA, $n_{IPA}$ = 1.369[23]) – and measuring responsivities for incident light in the wavelength range from λ = 1100 to 1600 nm. Devices were characterized under Incident light supplied by a glass fiber, i.e. without using optical instrumentation such as microscopes.

**Results and discussion**

Extraordinary optical transmission[24] through the NHA at the resonance wavelength in the presence of a liquid leads to peaks in responsivity in the measured (Fig. 2a) and simulated (Fig. 2b) spectra. This asymmetry of the peak line shape, which is more pronounced in the simulated spectra, is characteristic for a Fano resonance[25], and the resonance shows a clear shift as a function of the RI change induced by submerging the device in DI, EtOH or IPA. Experimental nonidealities such as roughness of the metal surface, nonplanarity of the metallization resulting from the fabrication process, as well as illumination that deviates from perfect vertical incidence, contribute to the peak broadening in the measured spectra. For a reference device without a NHA, the responsivity spectra are only weakly influenced by the superstrate refractive index (Fig. S1b). All measured spectra show a characteristic drop in responsivity at around λ = 1550 nm, corresponding to the direct transition energy of 0.8 eV in Ge. Finally, the dip at around λ = 1450 nm in the measured responsivity of the device submerged in DI water originates from an absorption band in water; this feature is not reproduced in simulation. Variation of the array pitch (Fig. 3a) leads to a shift in the resonance peak of the Fano resonance allowing to freely adjust the wavelength of operation within a wavelength range up to ~ 1.5 µm. The diameter of the holes is found to have a minor influence on the peak position when keeping the array's pitch constant (Fig. 3b). The Fano resonance gets broadened with increasing hole diameter due to increased background transmission through the array. Peak positions of the measured and simulated spectra for all array pitches are in good quantitative agreement (Fig. 2c). This confirms that the measured responsivity spectra are well reproduced by our simulations (finite-difference time-domain method, FDTD). The NHA pitch can be seen to strongly influence the peak wavelength, i.e. the operating point of the sensor.

The gradient between the points in Fig. 2c indicates the sensitivity $S$, which is a widely used quantity to evaluate the devices' suitability for RI sensing, defined as[26]

$$S = \frac{\Delta\lambda_R}{\Delta n_{Sup}}, \text{Eq. 1}$$

where $\Delta\lambda_R$ denotes the shift in resonance peak wavelength. When comparing the sensitivities for the two measured shifts in superstrate RI ($\Delta n_{Sup,1} = n_{DI} - n_{EtOH}$ and $\Delta n_{Sup,2} = n_{EtOH} - n_{IPA}$) for all characterized devices (Fig. 4a), $S$ can be seen to be largely independent of hole diameter. While the diameter has an influence on the resonance shape it does not affect its peak position (Fig. 3b). The sensitivities for a fixed pitch averaged over all nanohole diameters (horizontal lines in Fig. 4a) can be seen to vary with $\Lambda$ and with $\Delta n_{Sup}$. Devices with $\Lambda$ = 950 nm exhibit $S$ = 775 nm RIU$^{-1}$ on average for the index shift $\Delta n_{Sup,2}$ independently of the diameter of the holes. For the index shift $\Delta n_{Sup,1}$ the average sensitivity of devices with $\Lambda$ = 1000 nm is even higher (1180 nm RIU$^{-1}$). A comparison of our experimental results with experimental data from selected conventional, non-integrated plasmonic structures for refractive index sensing shows that the sensitivity in our setup is markedly higher than what can be observed for plasmonic Au nanoparticles[12] and, in particular, exceeds the sensitivities typically obtained in Al nanohole arrays[27] (see table 1).

*Table 1: Maximum sensitivities obtained experimentally for selected optical biosensor concepts based on plasmonic nanostructures.*

| Plasmonic nanostructure | Resonant wavelength (nm) | S (nm/RIU) |
|---|---|---|
| Au nanoprism[28] | 700 | 647 |
| Au nanocross[29] | 1400 | 500 |
| Au nanopillar[30] | 1300 – 1500 | 1000 |
| Periodic array of Au mushrooms[31] | 1260 | 1010 |
| Hybrid Au nanohole array[15] | 650 – 700 | 670 |
| Suspended NHA in Au film[32] | 795 | 717 |
| Passivated Al NHA[27] | 700 – 750 | 487 |
| This work | 1310 – 1350 | 1180 |

When the device is incorporated into a microfluidic setup with a laser diode as the vertical illumination light source[33], the operating wavelength $\lambda_{op}$ is fixed. In this operating mode, a resonance peak shift induced by a change in $n_{Sup}$ can be detected as a change in responsivity at $\lambda_{op}$. A dimensionless figure of merit FOM* to benchmark sensors with respect to signal amplitude rather than peak position change was first introduced by Becker et al.[34], and is defined as follows:

$$\text{FOM}^* = \max\left(\left.\frac{\frac{\Delta R_{\text{opt}}}{\Delta n_{\text{Sup}}}}{R_{\text{opt}}}\right|_\lambda\right), \text{Eq. 2}$$

Again, we determine FOM* by calculating $\frac{\frac{\Delta R_{\text{opt}}}{\Delta n_{\text{Sup}}}}{R_{\text{opt}}}$ as a function of wavelength for each RI change ($\Delta n_{\text{Sup},1}$ = $n_{\text{DI}} - n_{\text{EtOH}}$ and $\Delta n_{\text{Sup},2} = n_{\text{EtOH}} - n_{\text{IPA}}$) separately and extracting the maximum FOM* at the peak wavelength $\lambda_{\text{op}}$ (see Fig. S5). When comparing FOM* for all devices, we find a trend towards larger FOM* for smaller hole diameters (Fig. 4b). Reducing the hole diameter at a fixed pitch leads to a narrowing of the resonance peak, which leads to an increase in FOM*. Interestingly, throughout the wavelength range only a minor dependency of $\lambda_{\text{op}}$ on $d_{\text{NH}}$ was observed. The highest FOM* attained in all experimentally characterized spectra is 16. In comparison, optimized gold nanorods for plasmonic biosensing showed a FOM* around 24[34]. The values for FOM* extracted from the simulated spectra, however, are in excess of 100 and, thus, significantly higher (see Fig. S10). Indeed, the width of the simulated resonance peaks is considerably smaller, and the increase in responsivity between dip and resonance peak is steeper than in the measured spectra, which results in a larger FOM*. Thus, reducing structural variations introduced in the fabrication process can be predicted to have large potential for increasing the FOM* obtainable in our devices.

Thus, FOM* in particular depends crucially on the shape of the Fano resonance, whose origins were investigated in simulation. Our results indicate that the Fano resonance originates from a surface plasmon polariton (SPP) mode on the superstrate side of the nanohole array (Fig. 6b). SPPs can be excited on the metal-air interface of the square nanohole lattice by light at normal incidence and at wavelengths that fulfill the two-dimensional grating coupling condition

$$\lambda_{\text{SPP}}(i,j) = \frac{\Lambda}{\sqrt{i^2 + j^2}} \sqrt{\frac{\varepsilon_{\text{Al}} \varepsilon_{\text{Sup}}}{\varepsilon_{\text{Al}} + \varepsilon_{\text{Sup}}}}, \text{Eq. 3}$$

where *i* and *j* are integer numbers that denote the SPP mode order, and $\varepsilon_{Al}$ ($\varepsilon_{Sup}$) is the permittivity of Al (superstrate). A simulation of the electric field component $E_z$ normal to the NHA surface at the peak (λ = 1310 nm, Fig. 6b) shows that the electric field intensity decreases strongly with distance from the NHA. At the same time, a plot of the electric field intensity distribution along the *x-y* plane at the Al/superstrate interface reveals a propagating mode as is characteristic for SPPs – supporting the assumption. Evaluating Eq. 3 for a device with Λ = 950 nm and with superstrate index $n_{DI}$ yields $\lambda_{SPP}(1,0)$ = 1262.4 nm, which is close to the peak positions determined in simulation (λ = 1291 nm) and experiment (λ = 1315 nm).

While the measurement of sensitivities of our devices when subjected to bulk refractive index changes can be used to assess device performance, applications for biosensing require the accurate detection of refractive index changes in few nm thick surface biolayers. In biosensing applications, the metal Au is typically used, however, Au is difficult to integrate into a CMOS fabrication process. The CMOS compatible metal Al, which is used as diode metallization and for NHA in our devices, has also been explored for plasmonic applications[35]. To date there are few studies in which the surface functionalization of Al for the deposition of thin molecular layers has been analyzed[27,36]. Al forms a thin surface oxide film when exposed to air, whose properties can be controlled by exposing the metal to an oxygen plasma treatment, and which serves as a surface passivation that is able to protect the metal when exposed to water, oxygen and even diluted acids[27]. Suggested surface functionalizations for Al (or rather its surface oxide) include using organosilanes in combination with biotin−dextran−lipase conjugates[27] or polyethylene glycol (PEG)[36]. However, there is still lack of detailed knowledge about surface coverage ratio, film thickness and refractive index of the adsorbed biolayers, which makes it difficult to use these methods for sensor benchmarking purposes. In order to assess the suitability of our device for biosensing, we, therefore, simulate the refractive index changes induced by molecular binding events as follows: We model the molecular layer as a continuous film immersed in DI water with a fixed refractive index of $n_{film}$ = 1.45 and a saturation thickness $t_{film}$ = 6.5 nm (as determined for binding experiments of bovine serum albumin (BSA) and anti-BSA[37]). The resulting responsivity spectra as well as wavelength-dependent change in responsivity for two devices (Device 1 with Λ = 850 nm, $d_{NH}$ = 450 nm and Device 2 with Λ = 1000 nm, $d_{NH}$ = 450 nm) show that a responsivity change of up to ~ 0.015 A/W is predicted to occur (see Fig. 5b) when the NHA surface is fully covered by the biofilm. For incident laser power of 0.1 mW, this results in a change in photocurrent of 1.5 µA, which exceeds the measured diode dark currents (39 nA at external voltage $U$ = 0 V) by two orders of magnitude. We note that the shifts of the responsivity peak wavelength $\Delta\lambda_{film}$ obtained from simulations ($\Delta\lambda_{film}$ = 2.0 nm for Device 1 and $\Delta\lambda_{film}$ = 2.1 nm for Device 2) agree well with the approximation

$$\Delta\lambda_{film,approx} = S * (n_{film} - n_{DI})\left(1 - e^{-\frac{2t_{film}}{d_{tail}}}\right) = 2.7 \text{ nm}, \text{Eq. 4}$$

if we assume a depth $d_{tail}$ ~ 400 nm of the evanescent tail[38] (as extracted from simulation results for the electric fields, see Fig. 6b) and a bulk sensitivity of $S$ = 800 nm RIU$^{-1}$.

While an optimization of all relevant parameters for maximum sensitivity for a given refractive index change is outside of the scope of this manuscript (selected results for some of the parameters have been reported in Ref. 39), we nonetheless point out that adjusting thicknesses of metal layer, the SiO$_2$ layer and the semiconductor can potentially boost device performance (Fig. 6). The thickness of the Al layer in particular has a large influence on the peak responsivity and slope steepness attainable in our devices (Fig. 6, compare structures 1 and 3 as well as structures 2 and 4): with increasing Al layer thickness the coupling between surface waves on the two sides of the film is modified and it has been found previously that for metal film thicknesses below the skin depth, the transmission resonances become stronger with increasing film thickness[40]. Furthermore, we note that although the Fano resonance obtained in our devices involves a SPP mode on the superstrate side of the NHA, the layer structure of the substrate can nonetheless be shown to influence peak height and width of the Fano resonance. Increasing the SiO$_2$ layer thickness (Fig. 6, compare structures 2 and 5) underneath the Al layer lowers the substrate's effective RI, enabling a better matching of front and backside SPP modes and, as a consequence, an increase in transmission through the array. At the same time, the Fano resonance line shape becomes less pronounced. Using a bulk Ge substrate (Fig. 6, compare structures 1 and 2 as well as 3 and 4) reduces the overall performance of the device. This observation makes it seem likely that reflections from the underlying substrate are interacting with the plasmonic resonances and, as a consequence, directly alter the Fano resonance line shape[41].

Finally, our devices can be readily combined with a microfluidic setup consisting of microfluidic chips manufactured from cyclic olefin copolymer[34]. This fluidic chip material offers the possibility of large-scale fabrication by injection molding and can be used with our devices for CMOS-compatible biosensing, provided that a surface functionalization of the passivated Al surface specific to the targeted application can be found. While first results on the surface functionalization of passivated Al have been obtained[27,36], this aspect merits further investigation.

**Conclusion**

Our integrated collinear sensing system based on heterostructure Ge PIN photodetectors with Al nanohole arrays in their contact metallization uses Fano resonances to attain very high sensitivities > 1000 nm RIU$^{-1}$ in detecting bulk RI changes. The resonance wavelength can be adjusted by changing the nanohole array pitch in order to meet the application demands. Notably, only CMOS-compatible materials were used in sensor fabrication. Using a semiconductor heterostructure containing Ge is key to obtaining high photocurrents and eliminating the need for external instrumentation such as lock-in amplifiers.

Our simulation results agree well with the experimental data and offer insights as to the optimization of our setup for a chosen RI sensing scenario. We find that our device setup offers a large number of tuning parameters for optimization. In our setup in particular, the heterostructure composed of materials with different permittivities provides additional degrees of freedom that can be used to tune resonance shapes and, thus, the sensor response to RI changes. Given the difference in peak width between simulated and measured responsivity spectra, however, we note that the largest potential for improving the performance of our fabricated sensors lies in reducing fabrication imperfections such as local variations in metal thickness and thickness of the oxide layer covering the device. Our results show that structuring the NHA into the contact metallization of a Ge heterostructure photodetector can be used to fabricate a very compact, CMOS-compatible RI sensor that constitutes an important milestone towards utilizing the Si CMOS-platform for biosensors with plasmonic nanostructures in order to benefit from well-established large-scale production with high yields.

**Methods**

Using a solid phase molecular beam epitaxy system with co-evaporation of dopants the deposition of the layer stacks started with a 50 nm Si buffer on a 4 inch p$^-$-Si substrate. A 400 nm B-doped Si-layer ($N_A$ = 10$^{20}$ cm$^{-3}$) forms the p-doped region of the diode onto which an undoped Ge-layer with a thickness of $t_{Ge}$ – 20 nm was deposited, which was annealed afterwards at 900° C to reduce dislocations in the Ge-layer and enable the subsequent growth of high quality layers. The *n*-doped region contains a heterostructure with a 20 nm Sb-doped Ge-layer followed by a 20 nm Si-layer with $N_D$ = 10$^{20}$ cm$^{-3}$ for each.

Photolithography and reactive ion etching were used to define circular mesa structures with a diameter of 160 µm. After a cleaning step the structures were passivated with SiO$_2$ using a plasma-enhanced chemical vapor deposition process followed by a chemical mechanical polishing step to planarize the surface. This is necessary due to the metallization thickness of 100 nm which does not allow surface steps without the risk of a contact loss. A SiO$_2$ film with a thickness of 50 nm remained on top of the mesa. Contact holes were structured into the passivation layer by means of photolithography and reactive ion etching. The final structuring of the metallization was carried out by e-beam evaporation of 100 nm of Al followed by a lift-off process. The holes of the NHA were structured using PMMA-950k and EBL for masking within a subsequent etching process by reactive ion etching. The remaining EBL resist was removed by O$_2$ plasma ashing.

Devices were characterized under vertical illumination with optical power $\Phi_\lambda$ from the tip of a glass fiber placed above the NHA. The wavelength of the incident light was varied from 1100 nm to 1600 nm with a step size of 5 nm using a super continuum light source with an acousto-optical tunable filter (linewidth ~ 5 nm). The acquisition of one responsivity spectrum took ~ 5 minutes, the characterization of one device submerged in three different liquids took ~ 30 minutes. All devices were operated in photovoltaic mode at external voltage $U$ = 0 V. The wavelength-dependent change in responsivity under changes in $n_{Sup}$ was obtained by filling the space between glass fiber tip and device surface with deionized water (DI), Ethanol (EtOH) and Isopropanol (IPA). All liquids were of 'very large scale integration' (VLSI) quality.

The simulations were carried out using Lumerical's FDTD Solutions to obtain the electrical fields as well as the transmission through a unit cell of the structure as shown in Fig. 6c. The simulated responsivities of Fig. 2b were obtained by determining the losses occurring between a transmission monitor above and below the Ge layer (see Fig. 6c, structure 1 with 700 nm Ge instead of 500 nm). For detailed information we refer to chapter 3 of the supporting information.

**Associated content**

**Supporting information**

Measurement setup and device responsivity without NHA; influence of Ge layer thickness on the responsivity; procedure for peak wavelength extraction; FOM determination; dependence of Fano resonance and evaluation wavelength on hole diameter; details on the simulation setup and biofilm simulation.

**Author information**

**Corresponding author**

*E-Mail: lion.augel@iht.uni-stuttgart.de

**Notes**

The authors declare not to have any competing financial interests.

**Acknowledgements**

We acknowledge funding by the Ministry of Science, Research and Art Baden-Württemberg (MWK) and the University of Stuttgart through Research Seed Capital (RiSC). Funding period 2016 – 2019.

We thank the 1st Physics Institute of the University of Stuttgart for providing the permittivity data, especially Dr. Audrey Berrier.

**Figures**

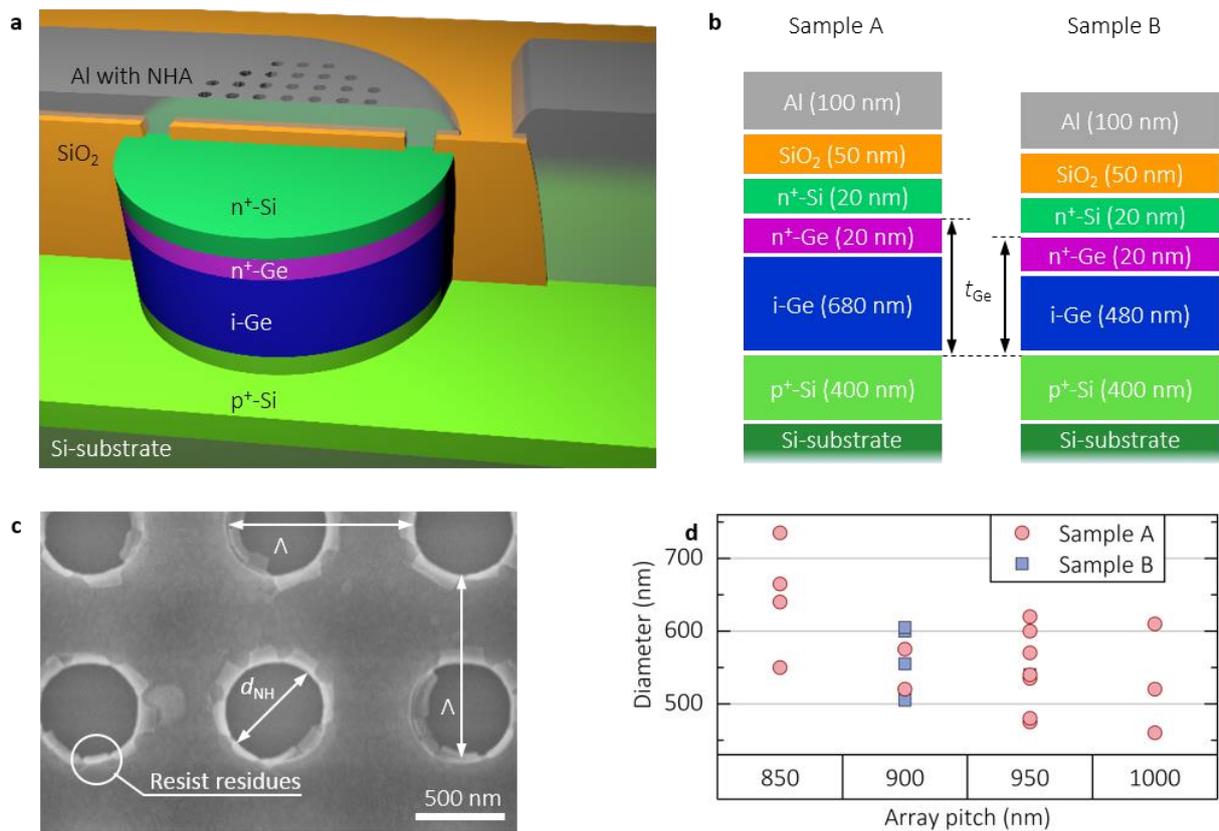

**Fig 1: Device design and NHA design parameters (a)** 3D schematic drawing of the Ge PIN photodiode with Al NHA within the contact metallization. **(b)** Schematic cross sections of the MBE layers fabricated for the use in the Ge PIN photodiodes with a total Ge layer thickness of $t_{Ge}$ = 700 nm (sample A) and $t_{Ge}$ = 500 nm (sample B). **(c)** SEM image of a NHA with hole diameter $d_{NH}$ and array pitch $\Lambda$ as indicated. Few resist residues remain from the EBL process. **(d)** Overview of all fabricated combinations of NHAs with nanohole diameter $d_{NH}$ and array pitch $\Lambda$ in the contact metallization of Ge PIN photodiodes.

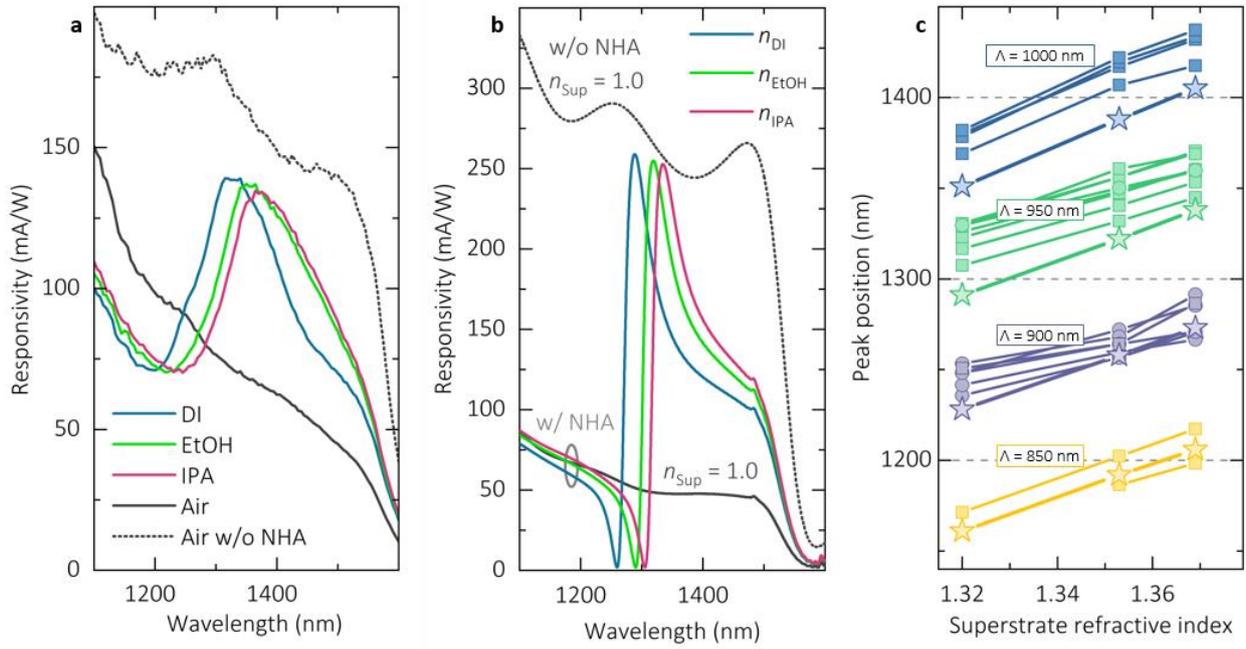

**Fig. 2: Optical device characteristics under the influence of a varying superstrate refractive index (a)** Measured responsivities for devices without and with a NHA with array pitch Λ = 950 nm and hole diameter $d_{NH}$ = 480 nm submerged in deionized water (DI), Ethanol (EtOH) and Isopropanol (IPA) to vary the superstrate refractive index. Both devices have a Ge layer thickness of $t_{Ge}$ = 700 nm. A change in superstrate refractive index causes a peak shift in the responsivities of the device with NHA. **(b)** FDTD simulation results for responsivities of devices with parameters identical to experiment, see (a), and superstrate refractive indices corresponding to DI, EtOH and IPA. The asymmetric Fano lineshape is more pronounced in simulation. **(c)** Peak positions $\lambda_R$ of the measured peaks in responsivities for all devices (○: $t_{Ge}$ = 500 nm, □: $t_{Ge}$ = 700 nm ) as function of the superstrate refractive index. Peak positions of the simulated responsivity spectra (star symbols, $t_{Ge}$ = 700 nm, $d_{NH}$ = 550 nm) show good agreement with experimental results.

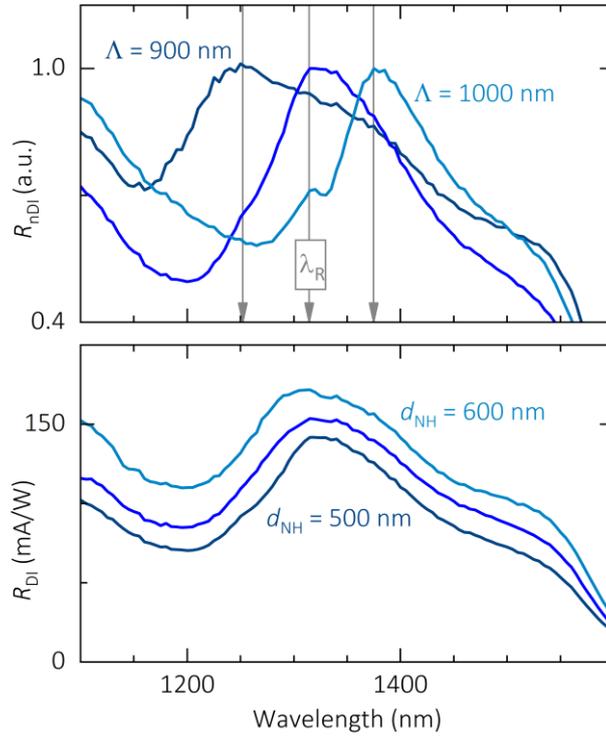

***Fig. 3: Influence of array pitch and hole diameter on device responsivity.** (a) Normalized responsivity for devices with different array pitch Λ. The responsivity spectra are been normalized to the corresponding peak values to facilitate the comparison of the spectra. (b) Responsivity spectra of devices with different hole diameters at Λ = 950 nm.*

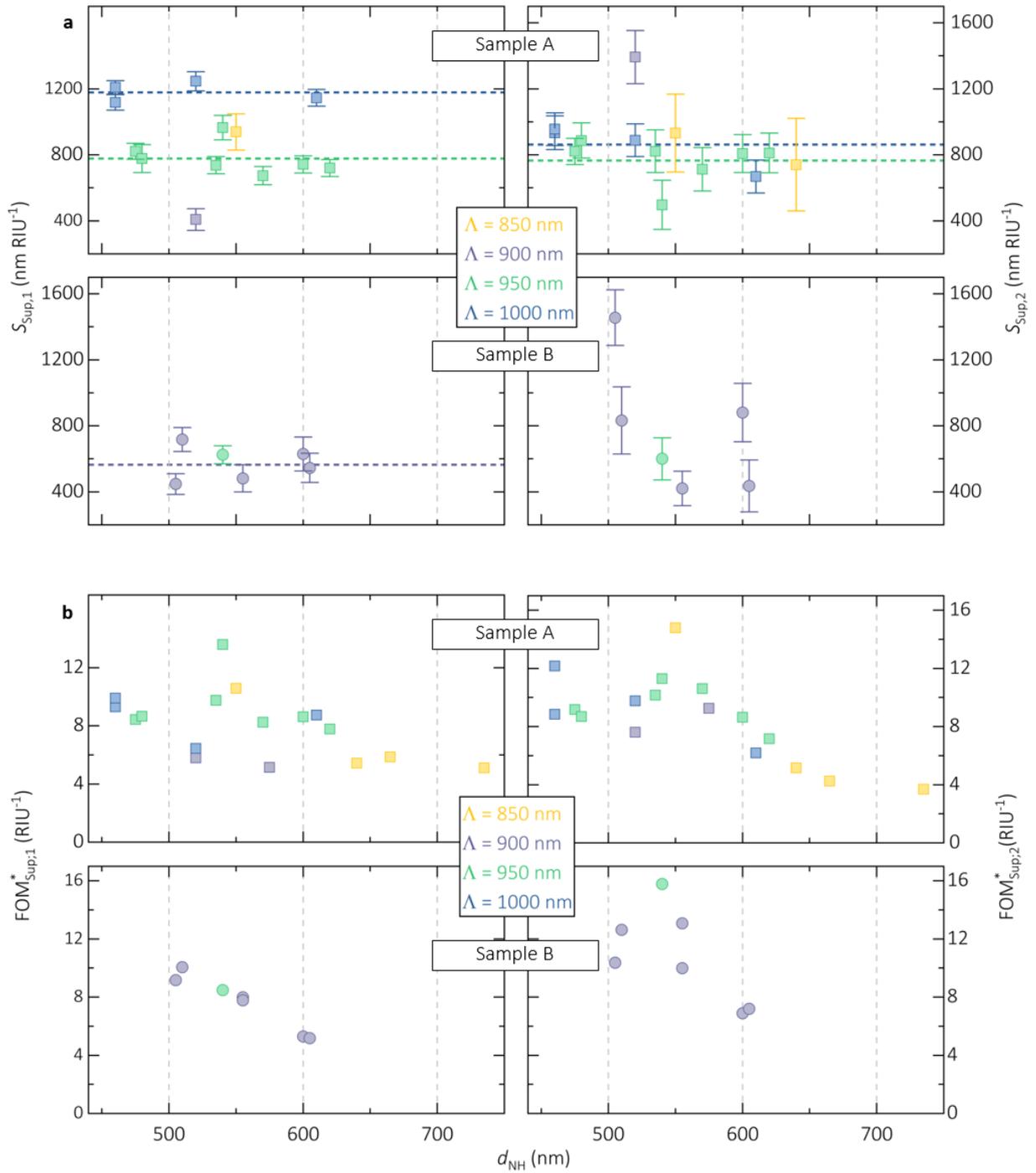

**Fig. 4: Figure of merits for all devices. (a)** Sensitivities extracted from peak fitting the responsivity spectra of all devices with errors. Sensitivities were extracted separately for the two measured shifts in superstrate refractive index ($S_{Sup;1}$ for the shift $\Delta n_{Sup,1} = n_{DI} - n_{EtOH}$ and $S_{Sup;2}$ for the shift $\Delta n_{Sup,2} = n_{EtOH} - n_{IPA}$). Dashed lines indicate the mean sensitivity for sets of devices with fixed array pitch. Mean values were only extracted for sets of devices with more than two different hole diameters at fixed array pitch. **(b)** FOM* extracted for all devices. Error bars were neglected since they are smaller than the symbol size. There is a trend towards increasing FOM* with decreasing hole diameter.

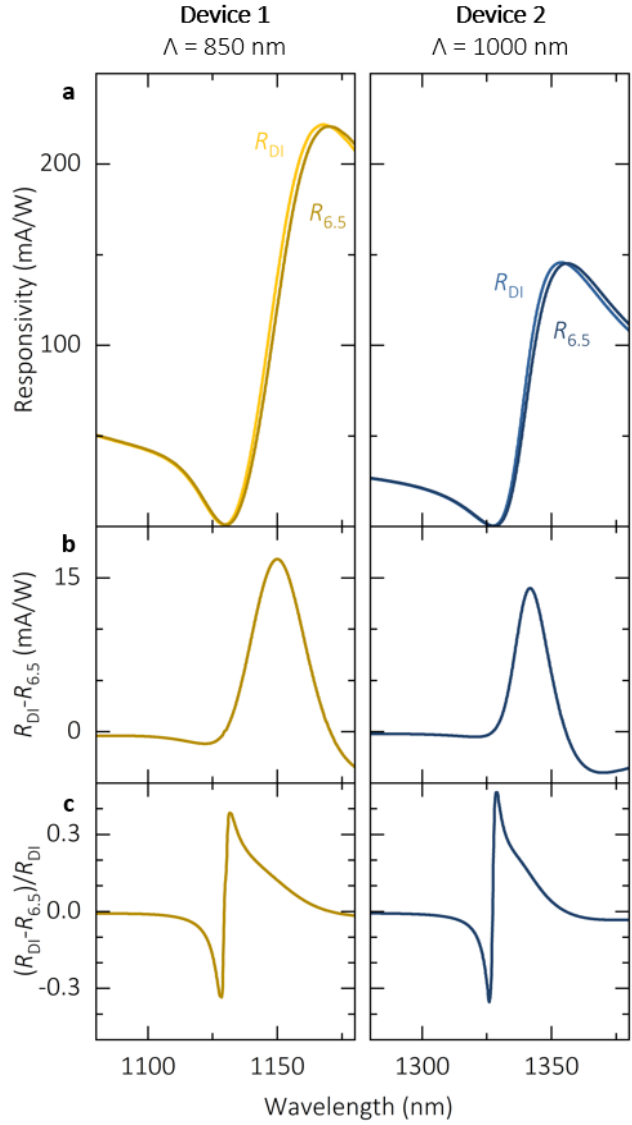

**Fig. 5: Simulated responsivities of two devices in the absence and the presence of a thin biofilm.** *(a)* Responsivity of two devices with Λ = 850 nm (Device 1) and Λ = 1000 nm (Device 2) before ($R_{DI}$) and after ($R_{6.5}$) the NHA surface is fully covered by a biofilm with a refractive index $n_{film}$ = 1.45 and a saturation thickness $t_{film}$ = 6.5 nm. *(b)* Responsivity change ($R_{DI} - R_{6.5}$) induced by the biofilm. *(c)* Responsivity change ($R_{DI} - R_{6.5}$) induced by the biofilm normalized to $R_{DI}$.

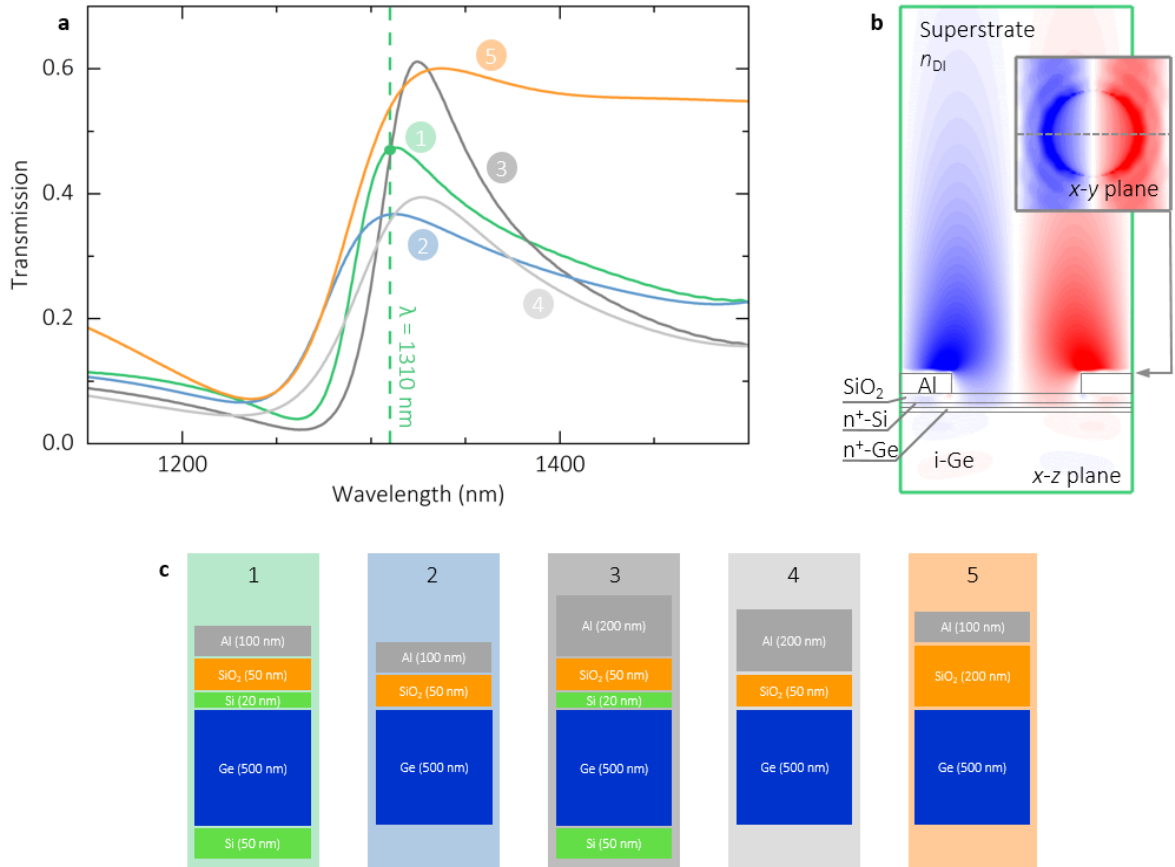

**Fig. 6: Simulated transmission spectra and electrical field distributions. (a)** Simulated transmission through different layer structures shown in (c) to illustrate the influence of the Si substrate, Al thickness $t_{Al}$ and oxide thickness $t_{SiO2}$ on the transmission through the NHA ($d_{NH}$ = 540 nm, $\Lambda$ = 950 nm). Structure 1 (green curve) corresponds to the experimentally investigated device layers. Both the width and the height of the transmission peak are influenced by the semiconductor layer structure as well as $t_{Al}$ and $t_{SiO2}$. **(b)** Simulation results for the z-component of the electrical fields in the x-z as well as x-y cross-sectional planes at the peak wavelength $\lambda$ = 1310 nm for structure 1. The dashed line within the inset indicates the position of the x-z cross-sectional plane. The simulation results reveal the presence of a SPP mode at the peak wavelength. **(c)** Schematic cross sections of the layer structures used for the simulations of transmission through a NHA in the topmost Al layer as shown in (a).

**SUPPORTING INFORMATION TO**

**Integrated collinear refractive index sensor with Ge PIN photodiodes**

Lion Augel, Yuma Kawaguchi, Stefan Bechler, Roman Körner, Jörg Schulze, Hironaga Uchida, Inga A. Fischer

The following pages contain:

# 1. Measured responsivity spectra

### a. Measurement setup and responsivity of devices with air as the superstrate

A schematic view of the measurement setup is given in Figure S1 (a). The diameter of the glass fiber was small (~ 9 μm) compared to the mesa diameter of the diodes (160 μm) to ensure that all incident light was coupled into the photodetector. By comparing the responsivities of the devices submerged in DI water (DI), ethanol (EtOH) or isopropanol (IPA) with the responsivity of the device in air it can be seen that the low refractive index of air shifts any resonances outside of the measured wavelength range.

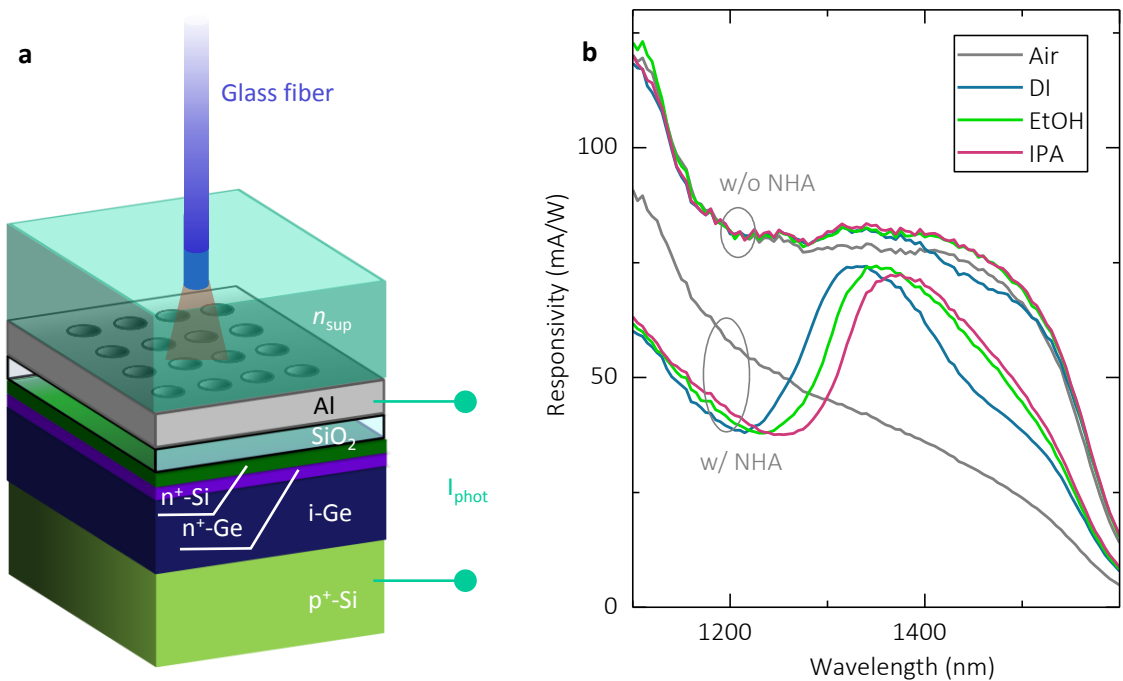

*Figure S1: (a) Schematic view of the illumination setup. A glass fiber is placed directly above the NHA. When the superstrate refractive index $n_{sup}$ is varied by immersing the devices in different liquids, the glass fiber is also immersed. (b) Plot of the responsivity with the three different fluids and in absence of any fluid on a device with and without a NHA. In absence of any fluid the Fano type line shape is not visible. Λ = 950 nm, $d_{NH}$ = 540 nm, $t_{Ge}$ = 500 nm.*

### b. Measured responsivity spectra for different Ge layer thicknesses, lattice pitches and NHA hole diameters

Increasing the *i*-Ge layer thickness (see *Figure S2*) leads to an increase in total responsivity.

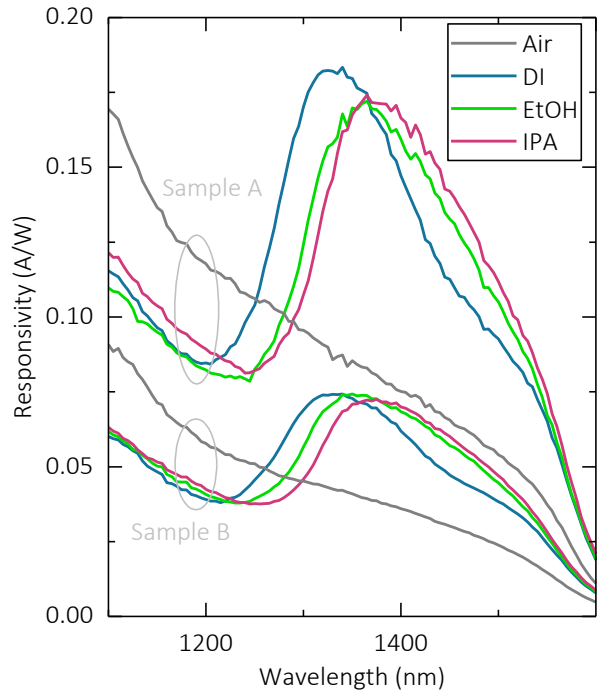

*Figure S2: Responsivities of two devices with i-Ge layer thicknesses of 480 nm (sample B) and 680 nm (sample A) and NHA parameters Λ = 950 nm and $d_{NH}$ = 540 nm.*

### c. Extraction of the peak wavelength

The determination of the sensitivities $S$ ($S = \frac{\Delta\lambda_R}{\Delta n_{Sup}}$, where $\Delta\lambda_R$ denotes the shift in resonance peak wavelength for a shift $\Delta n_{Sup}$ in superstrate RI) of the devices requires an extraction of the peak positions. Since our resonances are broadened by imperfections in materials and device fabrication, the peak shape can be expected to result from the convolution of a Fano line shape with Gaussian broadening. We achieve a good fit to the peak position as well as to spectral resonance shape at wavelengths below the peak wavelength by fitting the measured responsivity characteristics with Gaussian curves (*Figure S3*).

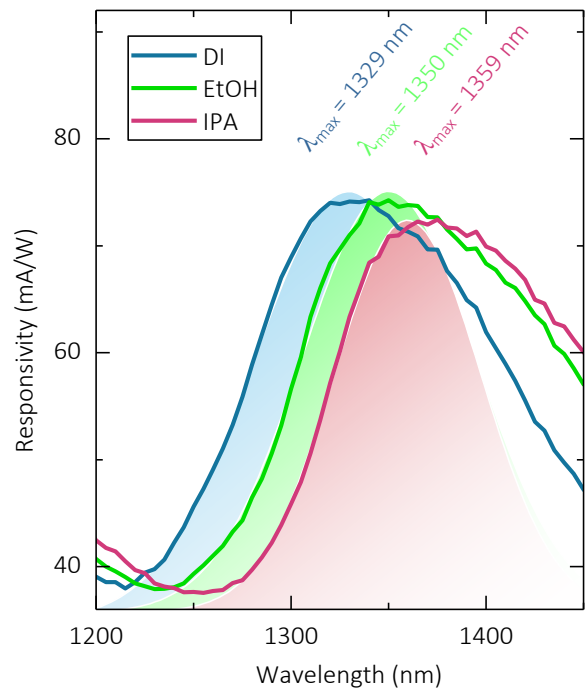

*Figure S3: Sample fits to the data using Gaussian curves to capture the broadening of the peaks.*

## 2. Sensitivity, FOM, and FOM*

### a. Dependence of peak wavelengths on hole diameter

The peak wavelength of our resonances depends strongly on the lattice constant Λ of the NHA but shows a negligible dependence on the hole diameter $d_{NH}$ as expected from a SPP resonance (*Figure S4*).

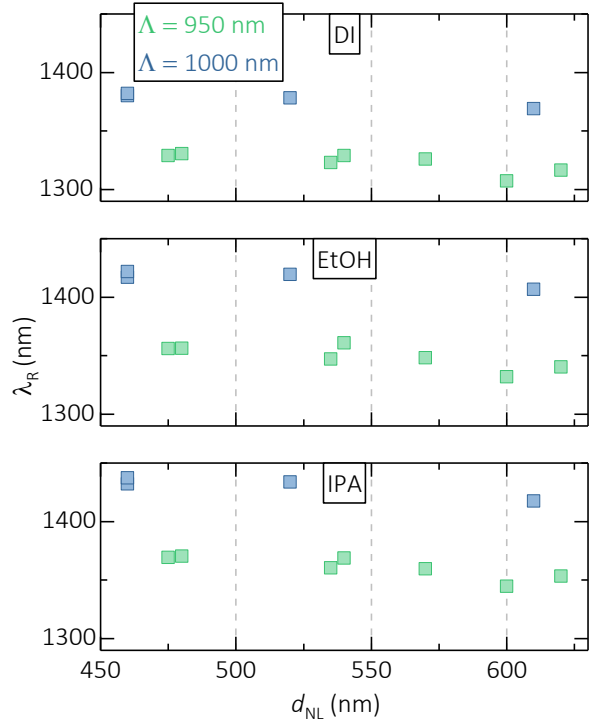

*Figure S4: Peak wavelengths extracted from the responsivity spectra. A slight drop in peak wavelength with increasing hole diameter can be seen as a result of peak widening.*

However, the resonance shape is influenced by the hole diameter, as a result, it is particularly important to verify that the evaluation wavelength used to extract FOM*, i.e. the wavelength at which

$$\frac{\frac{\Delta R_{opt}}{\Delta n_{Sup}}}{R_{opt}}$$

reaches its maximum, does not fluctuate or shift. Indeed, we found only a minor dependence of $\lambda_{op}$ on $d_{NH}$ (Figure 21). This is an important result as variations in fabrication can be assumed to have a negligible influence on the evaluation wavelength, which is set by an external light source.

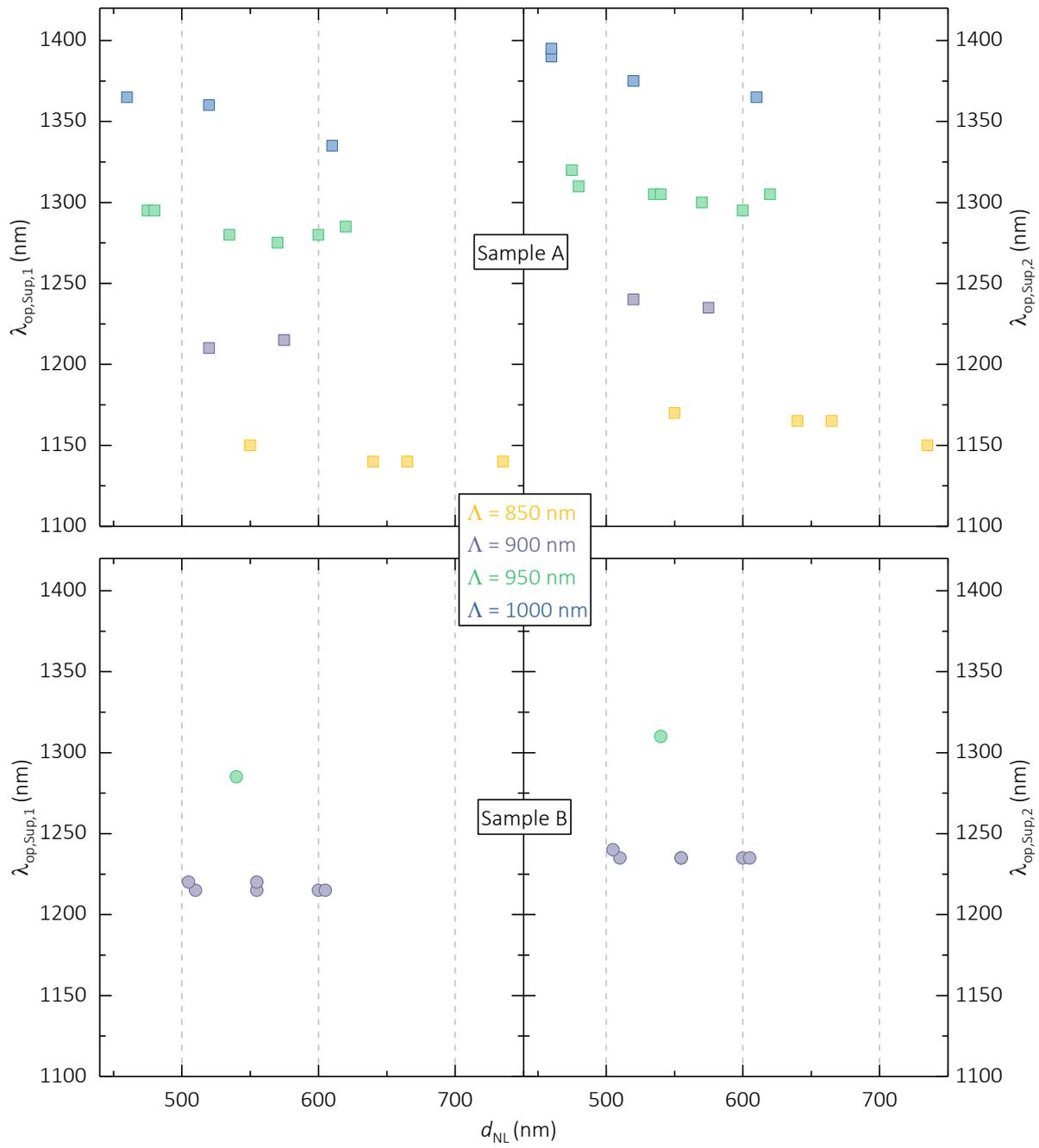

*Figure 21: The wavelength for the extraction of FOM* shows only weak variations with hole diameter.*

### b. Determination of FOM

When comparing different setups an additional figure of merit FOM is sometimes used:

$$\text{FOM} = \frac{S}{\Gamma_{\text{FWHM}}}$$

where $S$ is the sensitivity and $\Gamma_{\text{FWHM}}$ is the full width at half maximum of corresponding Gaussian fit (see figure S6). The values of FOM for our devices are given in Figure S6.

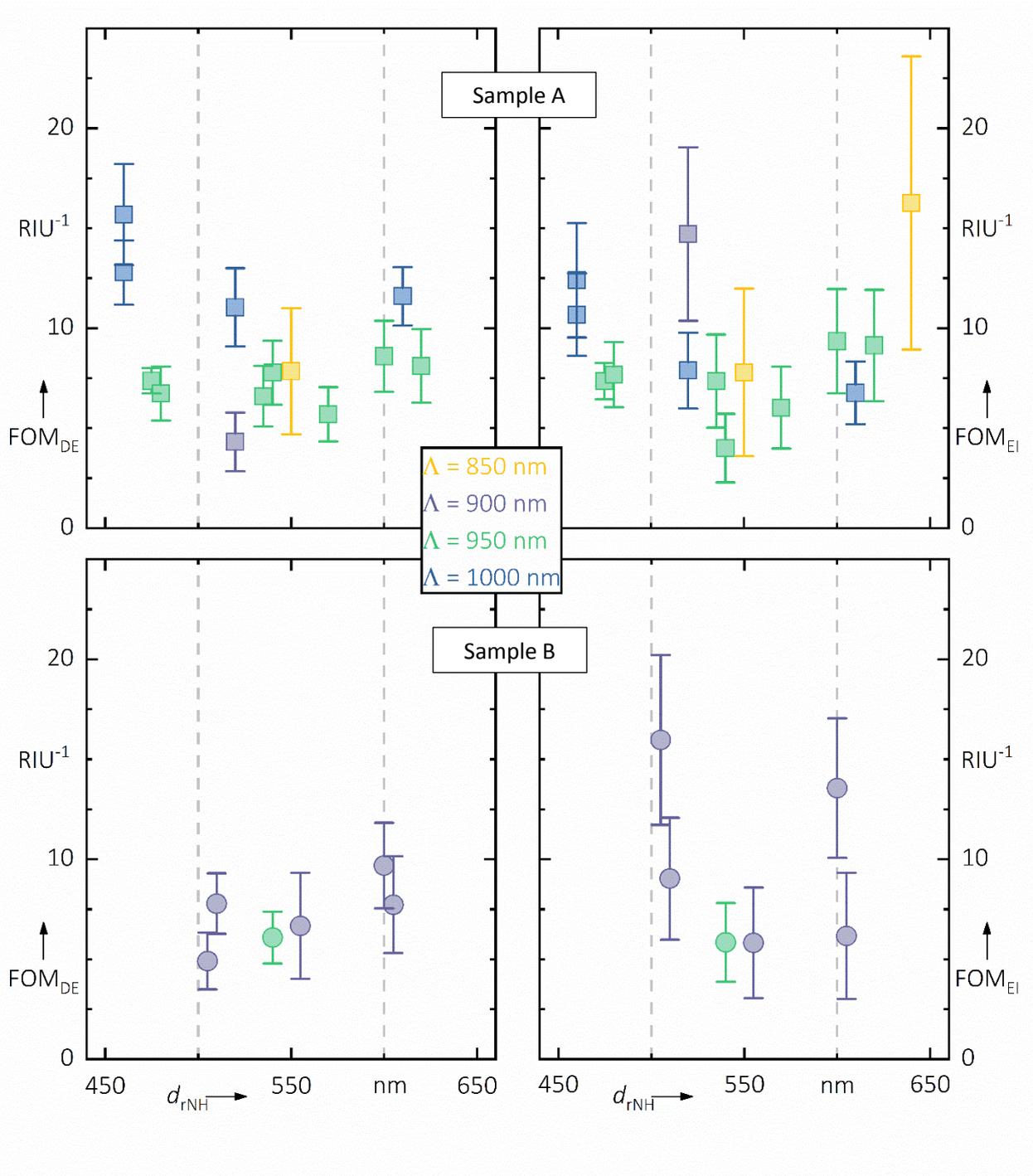

*Figure S6: FOM determined by Gaussian fits from the measured responsivities.*

## 3. Simulation

Electrical field profiles as well as transmission spectra were calculated using finite difference time-domain (FDTD) simulations in Lumerical's FDTD Solutions (www. lumerical.com). The modeled layer stacks from Fig. 6c (representing one unit cell) were arranged in a square array using Periodic Boundary Conditions (PBC) in *x* and *y* direction. In *z* direction Perfect Matched Layers (PML) extend the structure to infinity (see *Figure S7*). The incident transverse electromagnetic (TEM plane wave has a wavevector *k* normal to the surface of the layer stack. The mesh size of the structure was set to cubic voxels with 8 nm edge length.

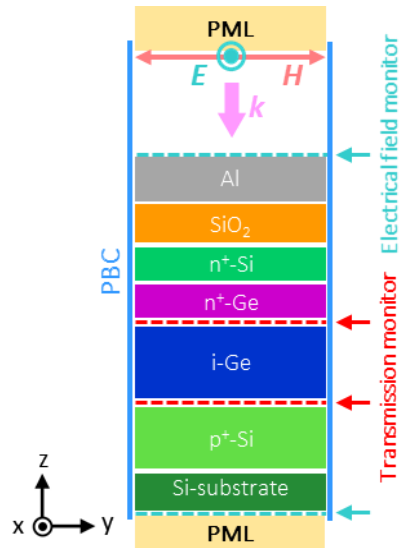

*Figure S7: Schematic view of the unit cell used for simulation*

The electrical field monitor was introduced at the metal-superstrate interface (*x-y* plane) as well as in the *x-z* plane cutting the hole exactly along the diameter. The transmission through the structure as well as the simulated responsivity were calculated from transmission monitors, calculating the transmission by

$$T = \frac{\frac{1}{2}\int Re(S_{plane})dA}{P_{source}},$$

where $T$ is the normalized power transmission, $S_{plane}$ is equal to the Poynting vector reaching the transmission monitor, $P_{source}$ is defined by electric and magnetic field injected by the source and $dA$ is the surface normal. In order to obtain the transmission spectra shown in Fig. 6a the monitor was placed underneath the layer stacks shown in Fig. 6c.

We assume that the responsivity $R$ can be calculated with the energy of photon using the expression:

$$R = \frac{\frac{1}{2}\int Re(S_{in})dA - \frac{1}{2}\int Re(S_{out})dA}{\frac{hc}{\lambda}} \frac{Q}{P_{source}}$$

$S_{in}$ and $S_{out}$ are the Poynting vectors obtained by two transmission monitors placed above and under the i-Ge layer. $h$ is Planck's constant, $c$ is the speed of light in free space, $\lambda$ is the wavelength and $Q$ is the elementary charge. All losses occurring in between the two monitors originate from the generation of electron-hole pairs (assuming a quantum efficiency of 1) defined by the imaginary part of the permittivity. Fabricated devices have quantum efficiencies < 1, thus, the experimentally determined responsivity is always lower than the simulated responsivity.

Furthermore, the responsivity spectra of the devices are well approximated by multiplying the NHA transmission spectrum with the responsivity spectrum of a reference device without a NHA (which would be the responsivity of a simple photodiode), i.e. resonances in the responsivity spectra can be studied by investigating the transmission through the NHA, taking the substrate into account via its refractive index (Figure S8).

The (complex) permittivity data for the investigated wavelength range was obtained by spectroscopic ellipsometry measurements performed on MBE layers identical to the ones used in device fabrication.

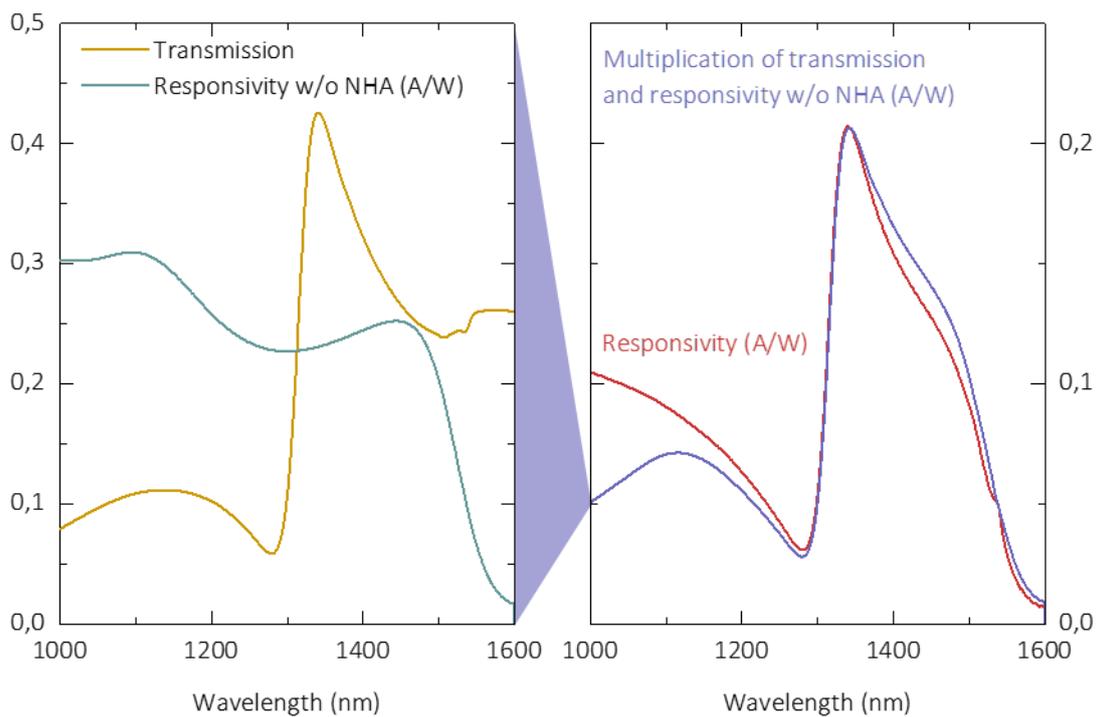

*Figure S8: The simulated device responsivity is well approximated by multiplying the transmission with the responsivity of a Ge PIN diode in the absence of a NHA, indicating that the peaks in the resonance can be understood by looking at the transmission only.*

### a. Determination of the superstrate's refractive index for simulation purposes

The superstrate's refractive indices or to be more precise the refractive index of the used fluids (DI, EtOH, IPA) were taken from literature (see main text for references). To reduce the effort in calculating the given figure of merits and the simulations, the refractive index was assumed to be constant over the investigated wavelength range and the value at λ = 1310 nm was used.

### b. Simulation results for the influence of parameter variations

Compared to simulation results, the measured responsivity spectra show a broadened peak shape as a result of a native oxide on the aluminum (Figure S9a), material inhomogeneity (compared to bulk aluminum), variations in $SiO_2$ thickness on the diode surface, and non-perpendicular angle of incidence (Figure S9b**Fehler! Verweisquelle konnte nicht gefunden werden.**,Figure S9c). The polarization angle was not found to have any influence on the device performance.

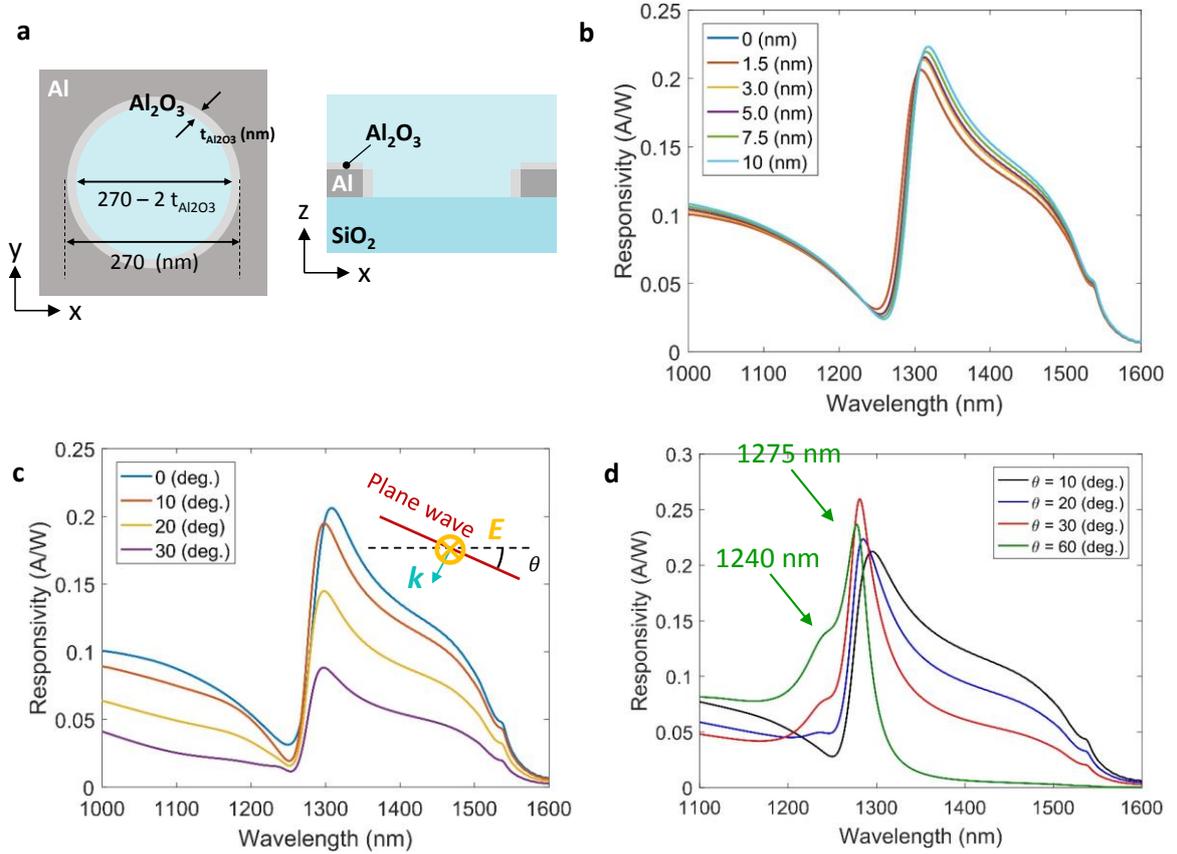

Figure 32: (a) Sketch of an $Al_2O_3$ layer with thickness varying from 0 to 10 nm on the NHA and (b) device responsivity for a NHA with $\Lambda$ = 950 nm, $d_{NL}$ = 540 nm, $t_{Ge}$ = 480 nm altered by the $Al_2O_3$ layer. (c) Device responsivity under s-polarized illumination with non-zero incident angle on a NHA with $\Lambda$ = 950 nm, $d_{NL}$ = 540 nm, $t_{Ge}$ = 480 nm. (d) Device responsivity under p-polarized illumination with non-zero incident angle on a NHA with $\Lambda$ = 950 nm, $d_{NL}$ = 540 nm, $t_{Ge}$ = 480 nm.

### c. Expected FOM*

From the simulated responsivities of devices with $d_{NH}$ = 540 nm, $t_{Ge}$ = of 480 nm, $t_{Al}$ = 100 nm and $t_{SiO2}$ = 50 nm, FOM* of above 100 were extracted as a function of lattice constant by the same method as used for the experimental data. The presence of a maximum should not be assumed since FOM* of smaller lattice constants is depending on simulation parameter (e.g. mesh size).

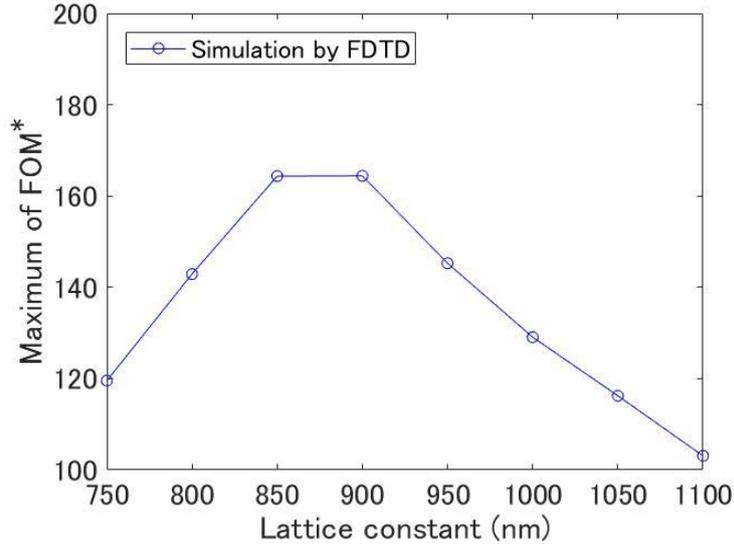

*Figure 33: FOM\* calculated from the simulation results.*

### d. Estimation of sensor response to molecular binding events

The device response on a molecular film has been modelled by a continuous and conformal film of varying thickness $t_{film}$ and a refractive index of $n_{film}$ = 1.45.

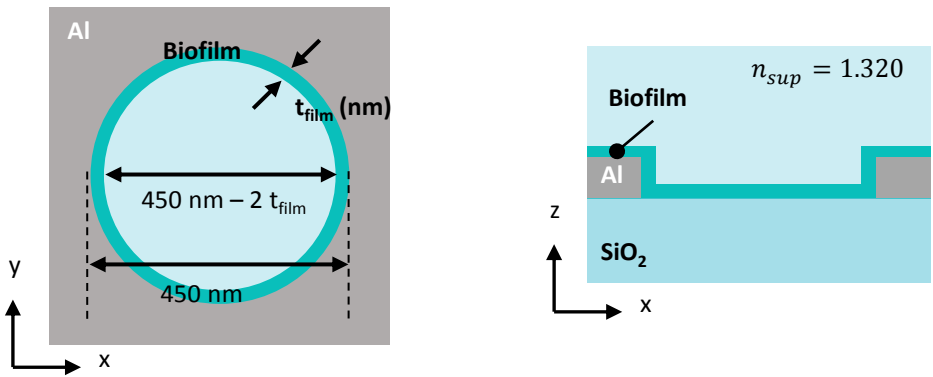

*Figure S11: Sketch of the upper part of the simulated device with molecular layer and the simulated responsivities. The simulated device parameters that are not mentioned in the sketch are: $t_{Ge}$ = 700 nm, $t_{Al}$ = 100 nm, $t_{SiO2}$ = 50 nm.*